\documentclass[%
 reprint,
 amsmath,amssymb,
 aps,
]{revtex4-2}

\usepackage{graphicx}
\usepackage{xcolor}
\usepackage{dcolumn}
\usepackage{bm}
\usepackage{braket}

\newcommand{\avg}[1]{\langle #1 \rangle}

\begin{document}

\preprint{APS/123-QED}

\title{Macro-level reinforcement tunes the transition order of reversible social contagion}

\author{Guanyu Zhang}
\affiliation{Department of Systems Science, Faculty of Arts and Sciences, Beijing Normal University, Zhuhai 519087, China}
\affiliation{International Academic Center of Complex Systems, Beijing Normal University, Zhuhai, 519087, China}

\author{Peng-Bi Cui}
\email{cuisir610@gmail.com}
\affiliation{Department of Systems Science, Faculty of Arts and Sciences, Beijing Normal University, Zhuhai 519087, China}
\affiliation{International Academic Center of Complex Systems, Beijing Normal University, Zhuhai, 519087, China}
\affiliation{School of Systems Science, Beijing Normal University, Beijing, 100875, China}


\begin{abstract}
Social contagion is often shaped by reinforcement: individuals become more likely to adopt a new behavior, opinion, or product as exposure accumulates or adoption becomes widely visible. Existing network models mainly capture this effect through local mechanisms, such as threshold responses or higher-order interactions. However, how macro-level reinforcement reshapes reversible spreading remains unclear. Here we study a SIS-like process in which pairwise transmission is reinforced by global prevalence. Combining quasistationary simulations and bifurcation analysis, we show that global feedback can produce first-order transition and hysteresis loop, with distinct activation and collapse thresholds. We further show how network localization promotes local ignition while weakening the global prevalence signal required for abrupt macroscopic activation, thereby raising the reinforcement threshold. Our results reveal how onset--retreat asymmetry emerges from global feedback coupled to network structure, providing a minimal mechanism for abrupt, history-dependent reversible social contagion.
\end{abstract}

\maketitle

\section{Introduction}\label{sec:intro}

Many social phenomena can be viewed as contagion processes on networks, including innovation diffusion, opinion spreading, and behavioral adoption~\cite{castellano2009statistical,valente1996network,iacopini2018network,kreindler2014rapid}. Classical compartmental models such as SIS and SIR provide a natural baseline for describing such spreading dynamics~\cite{PastorSatorras2015}. In their standard form, however, these models assume memoryless pairwise transmission with constant rates, leaving nonlinear adoption responses outside the description~\cite{centola2018experimental,gomez2016explosive,FerrazDeArruda2023,malizia2025hyperedge,su2017emergence}.

Social reinforcement provides a natural mechanism for such nonlinear responses. Repeated exposures, confirmation from multiple sources, and visible widespread adoption can reduce uncertainty, increase perceived legitimacy, and change the perceived payoff of adoption~\cite{Centola2010,CentolaMacy2007,young2015evolution,dimaggio2012network,wan2025diffusion}. Many existing models implement this effect locally, through threshold rules, repeated contacts, simplicial contagion, or hypergraph-based group interactions~\cite{granovetter1978threshold,Watts2002,tuzon2018continuous,Iacopini2019,li2021contagion,st2021social,StOnge2022}. These mechanisms have clarified how reinforcement can generate critical-mass effects, abrupt transitions, and bistability. Yet many social contagion processes are also shaped by population-level signals~\cite{onnela2010spontaneous,ma2014consumer,duan2009informational,bale2013harnessing}. The global prevalence of a behavior can affect its visibility, legitimacy, and perceived utility, especially in digitally connected systems where aggregate adoption is continuously displayed and amplified~\cite{bass1969new,katz1985network,markus1987critical,salganik2006experimental,muchnik2013social,denton2026conformity}. This motivates macro-level reinforcement, in which the global state of the population feeds back into individual-level transmission~\cite{Xue2026,xue2023identifying}.
   
Recent work incorporated such macro-level reinforcement into an irreversible SIR-like contagion process, showing that global feedback can transform a continuous outbreak transition into a mixed-order one, with the thresholds governed by network localization~\cite{Xue2026}. Many real spreading processes, however, are reversible rather than one-way. People may change opinions, abandon adopted technologies or platforms, and switch repeatedly between competing collective behaviors~\cite{castellano2009statistical,tandoc2019platform,hsiao2022network}. It remains unclear how macro-level reinforcement reshapes reversible spreading processes, and how network structure regulates the resulting dynamics.

Here we study a reversible SIS-like spreading process on networks, in which the pairwise transmission probability $\beta$ is reinforced by global prevalence~\cite{Xue2026,bass1969new,young2009innovation}. Combining quasistationary simulations~\cite{PastorSatorras2015,sander2016sampling} and bifurcation analysis, we show that macro-level reinforcement can turn the continuous SIS-like onset into a discontinuous first-order transition. The dynamics is organized by two thresholds: the intrinsic transmission threshold $\beta_c^{\rm high}$, at which the absorbing state loses stability, and the reinforcement threshold $\alpha_c$, above which the activation transition becomes discontinuous. When the reinforcement strength exceeds $\alpha_c$, the active branch persists below $\beta_c^{\rm high}$ and collapses only at a lower transmission probability $\beta_c^{\rm low}$, producing bistability and hysteresis. Thus, macro-level reinforcement allows activation and retreat to follow different dynamical paths in reversible social contagion.

We further show that network localization regulates these two organizing thresholds. Network localization allows activity to appear first around hubs, dense subgraphs, or other small structural cores, separating local ignition from macroscopic activation~\cite{Goltsev2012,Boguna2013,pastor2016distinct,st2021social,xue2023identifying}. This produces a threshold trade-off: localization lowers the threshold for local activation but raises the reinforcement strength required for abrupt finite-prevalence activation.

Taken together, these results extend macro-level reinforcement from irreversible outbreak dynamics to reversible, history-dependent contagion. They show that local ignition, abrupt macroscopic activation, and active-state persistence are governed by distinct dynamical conditions. This provides a minimal framework for understanding how population-level feedback and network structure jointly shape reversible social contagion, including platform adoption, technology switching, to opinion change. 

\section{Model}\label{sec:model}
We consider a reversible social contagion process on an undirected network with $N$ nodes, described by the adjacency matrix $\mathbf{A}$. Each node can be in one of two states: susceptible ($S$), meaning inactive or not currently adopting the contagion, and adopter ($A$), meaning currently active or adopting it. Let $Y_i(t)\in\{S,A\}$ denote the state of node $i$, and let $Y(t)$ denote the number of nodes in state $Y$ at discrete time $t$. In particular, $A(t)$ is the number of adopters at time $t$. At the beginning of update step $t$, the global adoption density is evaluated from the previous configuration as
\begin{equation}
    \rho(t) = \frac{A(t-1)}{N}.
\end{equation}
Adoption spreads through local pairwise contacts and is reinforced by this global adoption density. During the update from $t-1$ to $t$, each adopter attempts to activate each susceptible neighbor independently with transmissibility
\begin{equation}
    \beta'(t) = \min \bigl(1, \beta + \alpha\rho(t) \bigr),\quad \alpha \ge 0, \beta\in\left[0, 1\right].
\end{equation}
The parameter $\beta$ is the baseline pairwise transmissibility, characterizing the intrinsic attractiveness of the contagion, while $\alpha$ controls the strength of macro-level reinforcement. When $\alpha=0$, the model reduces to the standard discrete-time SIS model. For $\alpha>0$, a higher global density of adopters increases the likelihood of adoption, representing the effect of population-level signals such as visibility, perceived legitimacy, network utility, or collective pressure.

Adopter deactivate and return to the susceptible state with probability $\mu$ per time step. The contagion process can be written as:
\begin{equation}
\begin{aligned}
    S_i(t-1) + A_j(t-1) &\xrightarrow{\beta'(t)} A_i(t) + A_j(t), \\
    A_i(t-1) &\xrightarrow{\mu} S_i(t).
\end{aligned}
\end{equation}
All transition probabilities in update step $t$ are evaluated from the configuration at time $t-1$, and all state changes are then applied synchronously. We do not allow one-step reinfection: an adopter that deactivates during the update from $t-1$ to $t$ cannot be reactivated again within the same time step.

The linear feedback form is adopted as a minimal representation of macro-level reinforcement. It captures the leading-order effect of global adoption on local transmission while keeping the model analytically tractable. Without loss of generality, we fix $\mu=0.6$ in all simulations and MMCA calculations reported below.

\section{Results}\label{sec:results}
\subsection{Phase transitions and hysteresis}\label{subsec:phase_transition}
\begin{figure*}
    \centering
    \includegraphics[width=\textwidth]{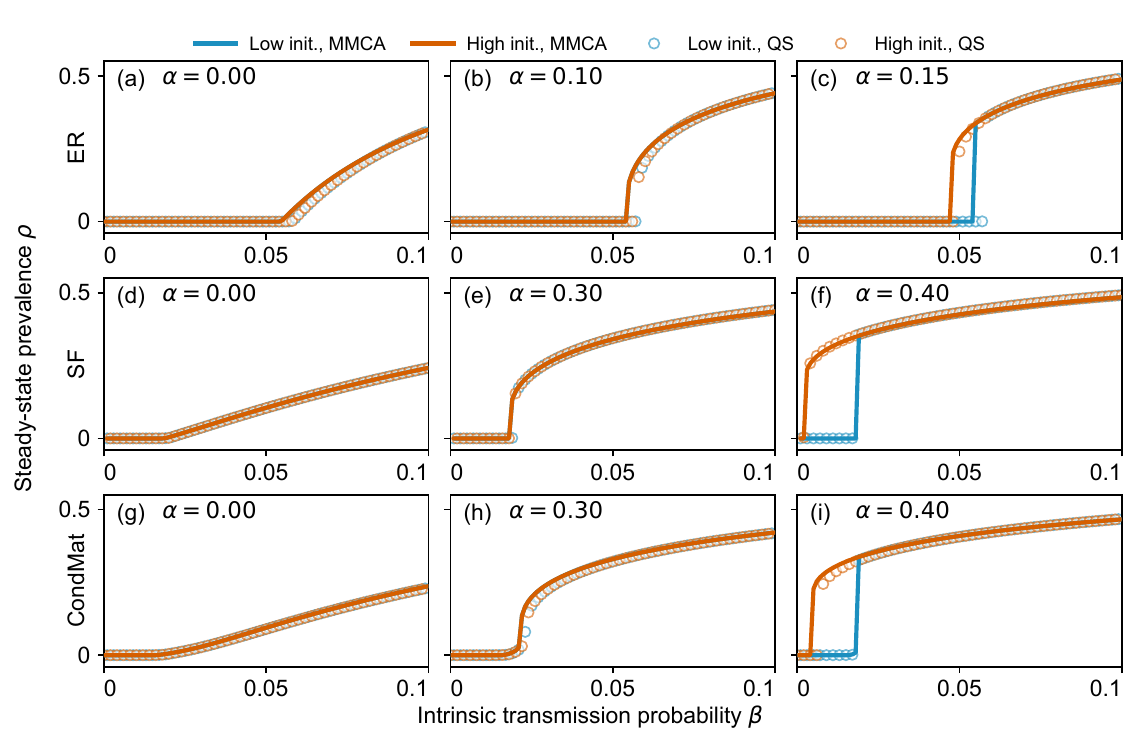}
    \caption{
    \textbf{Macro-level reinforcement induces abrupt transitions and hysteresis.} Steady-state prevalence $\rho$ as a function of the intrinsic transmission probability $\beta$ for ER (a--c), SF (d--f), and CondMat (g--i) networks at different reinforcement strength $\alpha$. Solid curves show MMCA stationary solutions initialized from low-prevalence and high-prevalence states, and open circles show the corresponding QS simulation results. When reinforcement strength is weak, the two initializations converge to the same branch and the onset remains continuous. When reinforcement strength is sufficiently strong, the branches separate: activation from low prevalence occurs through an abrupt jump, whereas retreat from the active state follows a distinct lower branch. The resulting loop indicates bistability and hysteresis, showing that macro-level reinforcement makes activation and collapse follow different dynamical paths.
    }
    \label{fig:fig1}
\end{figure*}
In finite SIS-like systems, activity cannot be created spontaneously, so the all-susceptible configuration is an absorbing state. Direct simulations therefore eventually fall into this absorbing state, even when a long-lived active regime exists. To sample the active quasistationary regime, we use quasistationary (QS) simulations, in which attempted visits to the absorbing state are replaced by previously visited active configurations. This approach allows absorbing-state transitions to be studied in finite systems and converges to the ordinary active-state behavior in the thermodynamic limit~\cite{dickman2002quasi,de2005simulate,ferreira2011quasistationary,sander2016sampling,costa2021simple}. In parallel, we use the microscopic Markov-chain approach (MMCA), a deterministic node-level description for the stationary activation probabilities, to track the corresponding stationary branches~\cite{arenas2023bifurcation}.

Using these two complementary approaches, we vary the intrinsic transmission probability $\beta$ at fixed reinforcement strength $\alpha$ for three different networks in Fig.~\ref{fig:fig1}, starting either from a low-prevalence state or from a high-prevalence active state. These two initial conditions probe whether the dynamics converges to a unique stationary branch or remains on distinct branches, thereby revealing possible history dependence.

Figure~\ref{fig:fig1} shows that weak reinforcement preserves the continuous SIS-like onset. The low- and high-prevalence initializations converge to the same branch, and the prevalence $\rho$ increases smoothly from the absorbing state. When reinforcement is sufficiently strong, the two branches separate. Starting from low prevalence, the system stays near the absorbing branch until it jumps abruptly to the active state. Starting from high prevalence, it follows an active branch that persists down to smaller values of $\beta$. The two transition points define a hysteresis window, indicating bistability between the absorbing and active states. Finite-size analysis in the Appendix~\ref{appendix:supp_transition_class} further confirms that this discontinuous transition has first-order character.

This behavior reflects the feedback between global prevalence and local transmission. Near the absorbing state, the prevalence $\rho$ is small, so the effective transmission probability $\beta+\alpha\rho$ is governed mainly by the intrinsic term $\beta$. Once a finite-prevalence state is established and the reinforcement strength $\alpha$ is large enough, the reinforcement term becomes appreciable and increases the effective transmission probability. This feedback stabilizes the active branch even when $\beta$ is reduced. Macro-level reinforcement therefore creates an asymmetry between activation and retreat in reversible social contagion.
\subsection{Bifurcation structure of reinforced spreading}\label{subsec:hysteresis}
\begin{figure*}
    \centering
    \includegraphics[width=\textwidth]{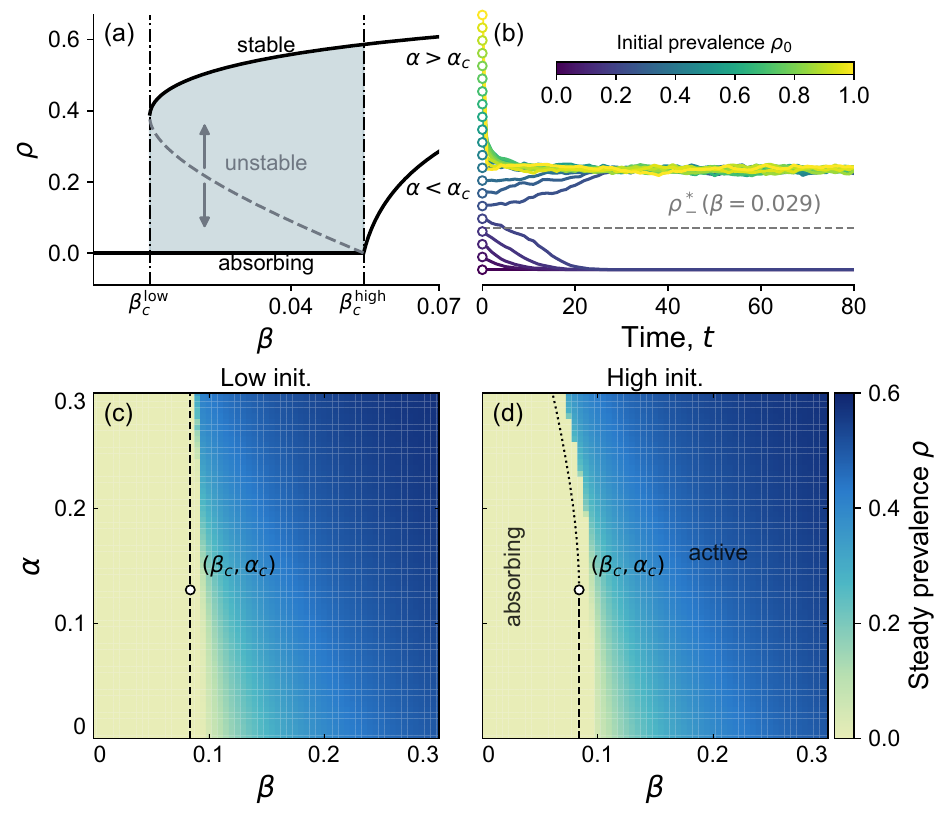}
    \caption{
    \textbf{Bifurcation structure of reversible spreading with macro-level reinforcement.}
    (a) Schematic bifurcation diagram from the reduced amplitude equation. For $\alpha<\alpha_c$, the active branch emerges continuously at the absorbing-state instability $\beta_c^{\rm high}$. For $\alpha>\alpha_c$, a stable active branch and an unstable saddle coexist below $\beta_c^{\rm high}$ and annihilate at the saddle-node boundary $\beta_c^{\rm low}$. (b) Time evolution from different initial conditions at fixed $\beta$ inside the bistable region. Initial conditions below the unstable saddle relax toward the absorbing branch, whereas those above it converge to the active branch. (c,d) QS phase diagrams initialized from low-prevalence and high-prevalence states. The dashed vertical line marks $\beta_c^{\rm high}$, the open circle marks $(\beta_c^{\rm high},\alpha_c)$, and the dotted curve in (d) marks the saddle-node boundary $\beta_c^{\rm low}(\alpha)$. The mismatch between the two initializations identifies the hysteretic region generated by the nonlinear sustaining branch.
    }
    \label{fig:fig2}
\end{figure*}
The reinforced MMCA defines a high-dimensional discrete-time dynamical system for the node activation probabilities. Let $\mathbf{p}=(p_1, p_2,\cdots, p_N)$, where $p_i$ denotes the probability that node $i$ is active. The absorbing state, $\mathbf{p}=\mathbf{0}$, is always a fixed point. The key question is how this fixed point loses stability and how macro-level reinforcement changes the nonlinear branch that appears near this instability.

To obtain an analytically tractable norm form, we expand the MMCA activation term to leading order in the activation probabilities of neighboring nodes. This gives the discrete-time quenched mean-field (QMF) equation
\begin{equation}
    \Delta p_i(t)= -\mu p_i(t) + \beta'(t)\bigl[1-p_i(t)\bigr]\sum_j a_{ij}p_j(t)
\end{equation}
where $\Delta p_i(t)=p_i(t+1)-p_i(t)$ and $a_{ij}$ is the element of the adjacency matrix $\mathbf{A}$. Near the absorbing-state threshold, the upper bound in $\beta'(t)$ is inactive, so $\beta'(t)=\beta+\alpha\rho(t)$.

Since the adjacency matrix $\mathbf{A}$ is real and symmetric, its eigenvalues can be ordered as $\Lambda_{\rm max}=\Lambda_1 \geq \Lambda_2\geq \cdots \geq \Lambda_N$. By the Perron--Frobenius theorem, the largest eigenvalues $\Lambda_1$ is positive and the corresponding normalized principal eigenvector (PEV) $\mathbf{u}=(u_1, u_2, \cdots, u_N)^T$ can be chosen non-negative~\cite{minc1988nonnegative,pillai2005perron}. Close to the absorbing-state threshold, the center eigenspace is spanned by the PEV, whereas the remaining modes are linearly stable and decay faster. We therefore write the activation-probability vector, to leading order, as
\begin{equation}
    p_i = x u_i + O(x^2),
\end{equation}
where $x$ is the amplitude of the PEV mode. Projecting the dynamics onto this critical mode gives the reduced amplitude equation:
\begin{equation}
    \Delta x = rx + cx^2 -dx^3,
    \label{eq:amplitude}
\end{equation}
with
\begin{equation}
    \begin{aligned}
        r &= \beta \Lambda_1-\mu,\\
        c &= \Lambda_1M_1(\alpha - \beta K),\\
        d &= \alpha \Lambda_1 M_1^2K.
    \end{aligned}
    \label{eq:rcd}
\end{equation}
Here $M_n = \sum_i u_i^n / N$ and $K = NM_3 / M_1$. The complete derivation is given in Appendix~\ref{appendix:derivation}. In this reduced equation, $r$ controls the linear stability of the absorbing state, the quadratic coefficient $c$ determines the direction of the bifurcating active branch, and the negative cubic term $-dx^3$ with $d>0$ provides nonlinear saturation.

The linear threshold is obtained from $r=0$~\cite{castellano2010thresholds,PastorSatorras2015,Goltsev2012}, yielding
\begin{equation}
    \beta_c^{\mathrm{high}} = \frac{\mu}{\Lambda_1}.
\end{equation}
This threshold is the point at which the absorbing state loses linear stability. Since the feedback contribution is proportional to the global prevalence $\rho$, it enters only beyond linear order and therefore does not appear in this linear intrinsic threshold. The character of the transition is instead controlled by the quadratic coefficient $c$. Setting $c=0$ at $\beta=\beta_c^{\rm high}$ gives the reinforcement threshold
\begin{equation}
    \alpha_c = \beta_c^{\mathrm{high}}K.
    \label{eq:theoretical_alpha_c}
\end{equation}
For $\alpha < \alpha_c$, the active branch emerges continuously from the absorbing state. For $\alpha > \alpha_c$, the nonlinear term reverses the direction of the branch: a stable finite-prevalence branch and an unstable saddle coexist below $\beta_c^{\mathrm{high}}$, producing bistability. The lower boundary of the hysteresis window is set by their saddle-node collision,
\begin{equation}
    \beta_c^{\mathrm{low}} = \beta_c^{\mathrm{high}} - \frac{(\sqrt{\alpha} - \sqrt{\alpha_c})^2}{K}.
\end{equation}
Thus the hysteresis width is
\begin{equation}
    \Delta \beta = \beta_c^{\mathrm{high}} - \beta_c^{\mathrm{low}} = \frac{(\sqrt{\alpha} - \sqrt{\alpha_c})^2}{K}.
\end{equation}
This bifurcation picture is summarized in Fig.~\ref{fig:fig2}(a). The absorbing state loses stability at $\beta_c^{\rm high}$, whereas the active branch disappears at $\beta_c^{\rm low}$. Between these two boundaries, the absorbing and active states coexist. The phase diagrams obtained from low- and high-prevalence initial conditions therefore reveal complementary basins of attraction: the former tracks the persistence of the absorbing state, whereas the latter tracks the stable active branch. Their mismatch gives the hysteresis region predicted by the saddle-node structure. This saddle-node mechanism provides the bifurcation-level origin of the discontinuous first-order transition confirmed by the finite-size analysis in the Appendix~\ref{appendix:supp_transition_class}.

The reduced theory also gives a dimensionless scaling form for the hysteresis width. Defining
\begin{equation}
        X = \frac{\alpha}{\alpha_c}, \quad Y = \frac{K\Delta \beta}{\alpha_c},
    \label{eq:collaspe_scaling}
\end{equation}
We obtain
\begin{equation}
    Y = (\sqrt{X} -1)^2.
    \label{eq:collaspe_curve}
\end{equation}
\begin{figure*}
    \centering
    \includegraphics[width=0.6\textwidth]{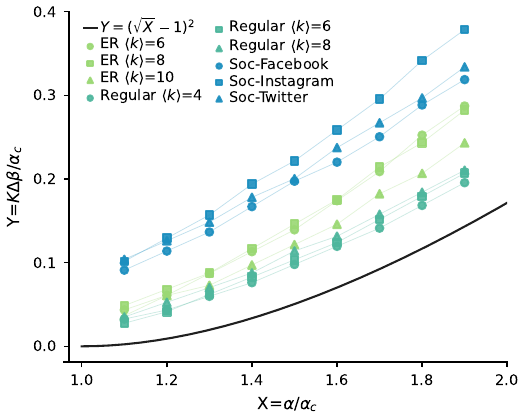}
    \caption{
    \textbf{Hysteresis windows are organized by the saddle-node scaling.}
    Dimensionless hysteresis width $Y=K\Delta\beta/\alpha_c$ as a function of normalized reinforcement strength $X=\alpha/\alpha_c$ for ER, regular, and empirical social networks. Symbols denote numerical measurements, and the solid black curve shows the saddle-node prediction $Y=(\sqrt{X}-1)^2$. The rescaled variables organize the widening of the hysteresis loop across different networks, indicating that the lower transition boundary is controlled by the saddle-node mechanism. Deviations from the theoretical curve reflect the main-mode and QMF approximations used in the reduced analysis.
    }
    \label{fig:fig3}
\end{figure*}
The rescaled data in Fig.~\ref{fig:fig3} follow this predicted nonlinear trend, showing that the widening of the hysteresis window is controlled by the same saddle-node mechanism.
\subsection{Network localization}\label{subsec:localization}
\begin{figure*}
    \centering
    \includegraphics[width=\textwidth]{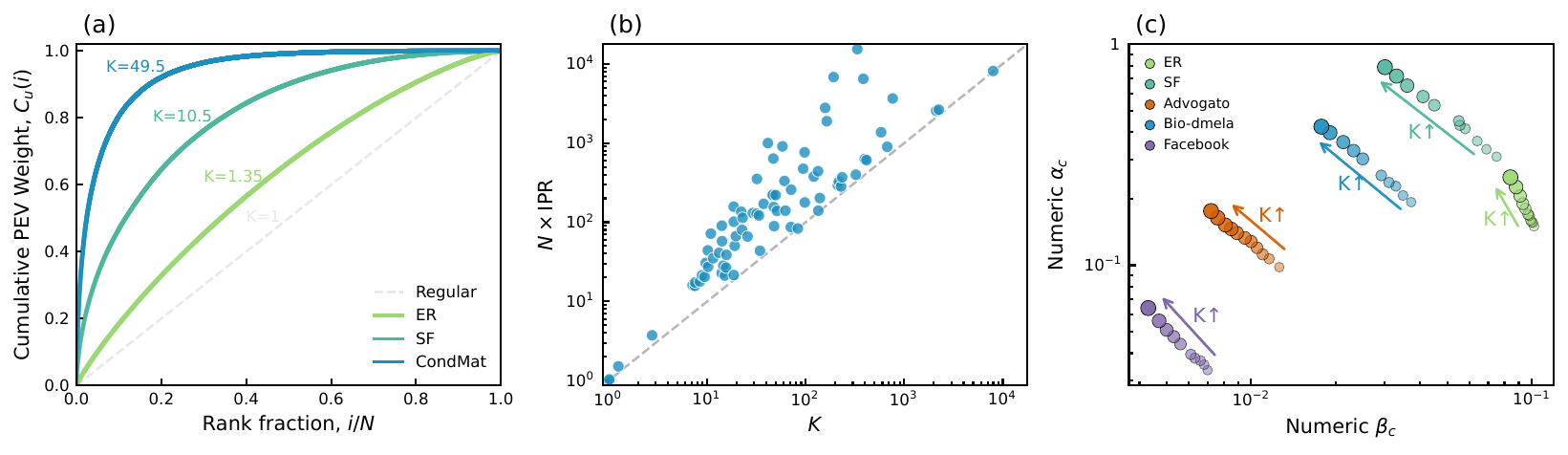}
    \caption{
    \textbf{Localization creates a trade-off between local activation and global reinforcement.}
    (a) Cumulative PEV weight $C_u(i)=\sum_{r=1}^{i}u_r^2$ as a function of the rank fraction $i/N$, with nodes ranked by decreasing PEV component. The dashed diagonal corresponds to a uniformly delocalized PEV, whereas upward-bending curves indicate stronger localization and larger $K$. (b) Relation between the localization coefficient $K=NM_3/M_1$ and $N\times{\rm IPR}$ across networks, confirming that $K$ captures the effective support of the PEV. (c) Numerical thresholds $\beta_c$ and $\alpha_c$ for degree-preserving rewired network families. Marker size increases along each rewiring trajectory, and arrows indicate increasing $K$. Stronger localization lowers the intrinsic threshold for local activation but raises the reinforcement strength required for abrupt finite-prevalence activation.
    }
    \label{fig:fig4}
\end{figure*}
The bifurcation analysis above is performed around the linear stability threshold $\beta_c^{\rm high}=\mu/\Lambda_1$. It describes the loss of stability of the absorbing state and the nonlinear branch generated along the PEV. The interpretation becomes especially important in localized networks. When the PEV is localized, activity just above $\beta_c^{\rm high}$ can remain concentrated on a finite number of nodes, giving a prevalence of order $\rho \sim O(1/N)$. Activity spreads beyond this localization center and involves a finite fraction of the network only at a higher value of $\beta$~\cite{Goltsev2012,Boguna2013,lee2013epidemic}. In contrast, when the PEV is delocalized, the instability at the spectral threshold directly produces a finite-prevalence active state.

Macro-level reinforcement makes this localization effect particularly important. The feedback term is generated by the global prevalence $\rho$, rather than the activity inside the localization center alone. A localized active state with $\rho=O(1/N)$ therefore produces only a vanishing reinforcement signal in the large-network limit. In this case, $\alpha_c=\beta_c^{\rm high}K$ should not be interpreted as the threshold for a macroscopic outbreak. Rather, it is the reinforcement strength required to reverse the character of quadratic coefficient at the linear threshold. The estimate of $\alpha_c$ can be vary large in strongly localized networks. Network localization thus separates the spectral onset of activity from the emergence of an endemic active state involving a finite fraction of nodes.

Interestingly, the coefficient $K=N\sum_i u_i^3/\sum_iu_i$ also provides a natural structural link between PEV localization and the two thresholds. If the PEV is effectively supported on $m$ nodes, with $u_i \sim m^{-1/2}$ on this support and negligible elsewhere, then $K\sim N/m$. Thus $K\sim O(1)$ for a delocalized PEV, whereas $K$ grows as the PEV weight becomes concentrated on a smaller subset of nodes. Fig.~\ref{fig:fig4}(a) visualizes this concentration by plotting the cumulative PEV weight against node rank. Curves close to the diagonal correspond to nearly uniform networks, while strongly upward-bending curves indicate that most of the PEV weight is carried by a small fraction of nodes. This interpretation is consistent with the inverse participation ratio, a standard measure of network localization,
\begin{equation}
    \mathrm{IPR} = \sum_i u_i^4.
\end{equation}
For an effective support of size $m$, one has $\mathrm{IPR} \sim 1/m$ and therefore $N\times\mathrm{IPR} \sim K$, as shown in Fig.~\ref{fig:fig4}(b).

Localization affects the two thresholds in opposite ways. A localized PEV is often associated with a hub core or dense substructure, which can become active at a smaller intrinsic transmission probability. This lowers the spectral activation threshold $\beta_c^{\rm high}$ and makes the network easier to ignite locally. However, the same localized activity contributes only weakly to the global prevalence $\rho$, which vanishes in the large-network limit. The network therefore requires a larger reinforcement strength $\alpha$ to turn local activity into a macroscopic active state and produce a global discontinuous transition. 

Fig.~\ref{fig:fig4}(c) tests this prediction using degree-preserving rewired network families. These networks are generated using the Xulvi-Brunet-Sokolov algorithm, which preserves the degree sequence while tuning the assortativity of networks, and hence the localization strength~\cite{xulvi2004reshuffling,Xue2026}. It should be noted that the assortativity does not strictly monotonically determine the strength of network localization. We use it here to reorganize edges and thereby change the localization structure. Along these rewiring trajectories, larger $K$ is associated with lower $\beta_c^{\rm high}$ but higher $\alpha_c$. Thus, localization creates a threshold trade-off: it facilitates local ignition, but makes an abrupt system-wide spreading event harder to trigger through macro-level reinforcement.
\begin{figure*}
    \centering
    \includegraphics[width=\textwidth]{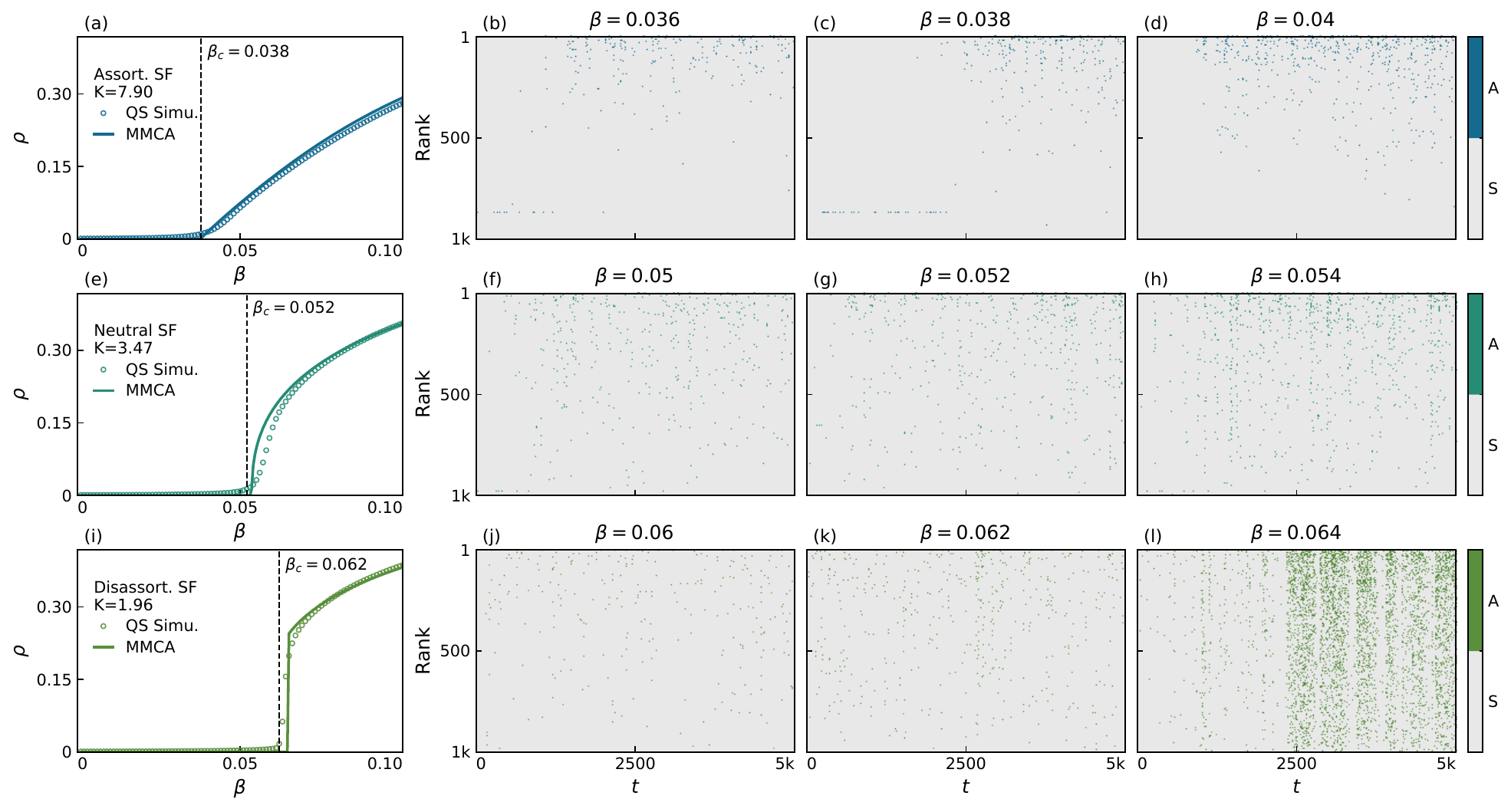}
    \caption{
    \textbf{Localization confines the escape from local activation to global spreading.}
    (a,e,i) Steady-state prevalence $\rho$ as a function of the intrinsic transmission probability $\beta$ at fixed reinforcement strength $\alpha=0.24$ for rewired SF networks with different localization strengths. Open circles show QS simulations, solid curves show MMCA predictions, and dashed vertical lines mark the numerical transition points. (b--d, f--h, j--l) QS time traces of node states below, near, and above the transition, with nodes ranked by decreasing PEV component. In strongly localized networks, activity near the transition remains concentrated on high-centrality nodes and produces only a weak global prevalence signal. As localization weakens, activity escapes more readily from the localized core and develops into a finite-prevalence active state.
    }
    \label{fig:fig5}
\end{figure*}

To examine how this trade-off occurs, we fix $\alpha$ and rewire a scale-free network, obtaining three networks with different localization strength $K$ and reinforcement threshold $\alpha_c$. As shown in Fig.~\ref{fig:fig5}, the same value of $\alpha$ can lie below, near, and above the reinforcement threshold of different networks. The observed transition therefore changes from a continuous onset to a discontinuous transition with hysteresis.

The time-resolved node states in Fig.~\ref{fig:fig5} reveal the microscopic path underlying these different transitions. Nodes are ranked by decreasing PEV weight, so small ranks correspond to the localization core. Near the activation threshold, activity first appears around these high-weight nodes. In the strongly localized network, activity remains confined near the localization core and spreads only gradually as $\beta$ increases. The prevalence therefore grows smoothly, so the feedback contribution $\alpha \rho$ is amplified only gradually. By contrast, in the weakly localized network, activity escapes the core more easily and reaches a broader set of lower-weight nodes. This produces a sharp increase in the global prevalence $\rho$. The increased prevalence raises the effective transmission probability $\beta+\alpha\rho$, which further amplifies spreading and produces a macroscopic jump. Thus, at fixed reinforcement strength $\alpha$, tuning the network localization changes whether localized activity remains confined and expands gradually, or is amplified into a global active state.

\section{Discussion}\label{sec:discussion}
A central question in social contagion is how microscopic transmission and network structure combine to produce macroscopic collective change. Our QS simulations and bifurcation analysis show that macro-level reinforcement does not merely increase the transmission probability. Instead, it reorganizes the stability structure of the contagion process. Compared with irreversible SIR-like dynamics with macro-level reinforcement~\cite{Xue2026}, the reversible SIS-like dynamics studied here contains a persistent active branch. Sufficiently strong global feedback creates a saddle-node structure below the absorbing-state threshold, so that activation from low prevalence and retreat from the active state are controlled by different stability boundaries. This produces bistability, hysteresis and a first-order transition, implying that growth and decline need not follow the same dynamical path. Activation and retreat are therefore intrinsically asymmetric processes.

Network localization determines whether this nonlinear mechanism remains confined or develops into a macroscopic active state. In localized networks, activity first concentrates around hubs, dense subgraphs, or other small structural cores, generating only a weak population-level signal. Strong localization therefore favors local ignition while suppressing the global feedback required for abrupt finite-prevalence activation. In delocalized networks, by contrast, activation may require a larger intrinsic transmission probability, but the emerging activity couples more efficiently to global prevalence. In this sense, localization acts as a bottleneck in converting local activity into global feedback: it confines early activity and weakens the global feedback.

This result complements existing studies of reinforced and complex contagion. Threshold, repeated-exposure, simplicial, and hypergraph models have shown that nonlinear adoption can produce critical-mass effects, bistability, and abrupt transitions~\cite{granovetter1978threshold,Watts2002,tuzon2018continuous,Iacopini2019,li2021contagion,st2021social,StOnge2022}. In these models, the nonlinear response is mainly encoded in local exposure or group interactions. However, in contemporary digital environments, popularity, prevalence, and other aggregate adoption signals are often visible to individuals, and their reinforcing effects on individual decisions should not be neglected~\cite{bass1969new,ma2014consumer,duan2009informational,bale2013harnessing,denton2026conformity}. Our results show how such macro-level reinforcement reshapes reversible spreading dynamics, and how network localization regulates its effect, providing a complementary perspective on reinforced diffusion in real-world systems.

Our model is intentionally minimal and maintains analytical tractability. The linear dependence of transmission on global prevalence should be viewed as a lowest-order approximation of population-level feedback, not as a universal empirical form. Real diffusion processes may involve nonlinear, delayed, heterogeneous, or saturating feedback, as well as temporal or adaptive contact structures~\cite{wang2015dynamics,holme2012temporal,vazquez2007impact,gross2006epidemic}. Nevertheless, the central physical picture should extend beyond the specific feedback function studied here: reversible reinforced spreading is governed by the competition between localized activation and global prevalence-mediated feedback. This framework provides a reference point for future studies incorporating higher-order, delayed, or heterogeneous forms of macro-level reinforcement.

\appendix

\section{NUMERICAL METHODS}\label{appendix:numerical_methods}
We use quasistationary (QS) simulations to sample the active steady state of the stochastic dynamics. In finite SIS-like systems, the all-susceptible configuration is absorbing, and direct simulations eventually fall into this state even when a long-lived active regime exists. In the QS procedure, each attempted visit to the absorbing configuration is replaced by an active configuration randomly selected from a stored set of previously visited active configurations. After discarding transients, observables are measured in the QS regime and averaged over time and independent realizations. This method is standard for stochastic processes with absorbing states and is designed to sample the quasistationary active regime directly.

For each network and parameter set, we measure the stationary prevalence $\rho$. The susceptibility is defined as
\begin{equation}
    \chi = N \frac{\avg{\rho^2} - \avg{\rho}^2}{\avg{\rho}},
\end{equation}
For $\alpha=0$, the activation threshold $\beta_c^{\mathrm{high}}$ is estimated from the peak of $\chi(\beta)$. For $\alpha > 0$, we perform continuation scans in $\beta$ from high-prevalence initial conditions, giving the lower boundary $\beta_c^{\mathrm{low}}$ in the same way. When $\alpha > \alpha_c$, the two branches identifies the hysteretic region, and the hysteresis width is defined as
\begin{equation}
    \Delta \beta = \beta_c^{\mathrm{high}} - \beta_c^{\mathrm{low}}.
\end{equation}
The reinforcement threshold $\alpha_c$ is determined from the onset of branch separation between low- and high-prevalence scans. The detailed numerical protocol used to estimate $\alpha_c$ is provided in the Appendix~\ref{appendix:numeric_alphac}.

\section{MICROSCOPIC MARKOV CHAIN APPROACH}\label{appendix:mmca}
We use the microscopic Markov chain approach (MMCA) as a deterministic node-level approximation to the stochastic spreading dynamics. Let $p_i(t)$ denote the probability that node $i$ is an adopter at time $t$, and let
\begin{equation}
    \rho(t) = \frac{1}{N}\sum_{i=1}^Np_i(t)
\end{equation}
be the corresponding global prevalence. 

Under the standard MMCA independence approximation, the probability that node $i$ is not activated by any neighbor at time $t$ is
\begin{equation}
    q_i(t) = \prod_{j=1}^N [1-\beta'(t)a_{ij}p_j(t)], 
    \label{eq:q_t}
\end{equation}
where $a_{ij}$ is the element of adjacency matrix $\mathbf{A}$. In this work, we use the no-reinfection version of the discrete-time dynamics, consistent with the stochastic simulation. Thus, an adopter that deactivates within a time step cannot be reactivated again in the same time step. The update equation is
\begin{equation}
    p_i(t+1) = [1 - p_i(t)][1-q_i(t)] + (1-\mu)p_i(t)
\label{eq:mmca}
\end{equation}
Together, these equations define a deterministic map
\begin{equation}
    \mathbf{p}(t+1) = \mathbf{F}(\mathbf{p}(t);\alpha,\beta,\mu),
\end{equation}
on the state space $[0, 1]^N$. The stationary MMCA prevalence is obtained by iterating this map from the given initial condition until convergence.

\section{DERIVATION OF THE BIFURCATION ANALYSIS}\label{appendix:derivation}
We derive the reduced bifurcation equation used in the main text. Starting from the no-reinfection MMCA update equation, we approximate the probability that node $i$ receives at least one successful activation attempt by retaining only single-neighbor activation events. This neglects overlap terms associated with simultaneous activation attempts from multiple neighbors, while keeping the susceptible factor $1-p_i$. Thus,
\begin{equation}
    1-q_i(t) \simeq \beta'(t)\sum_j a_{ij}p_j,
\end{equation}
where $a_{ij}$ is the element of adjacency matrix $\mathbf{A}$. Substituting this approximation into the MMCA update equation~\eqref{eq:mmca}, we obtain the discrete-time quenched mean-field (QMF) equation~\cite{PastorSatorras2015,de2018fundamentals,prasse2019viral}
\begin{equation}
    p_i(t+1)-p_i(t) = -\mu p_i(t) + \beta'(t)\left[1-p_i(t)\right]\sum_ja_{ij}p_j(t)
\end{equation}
Defining $\Delta \mathbf{p}(t) =\mathbf{p}(t+1)-\mathbf{p}(t)$, the equation can be written in vector form as
\begin{equation}
    \Delta \mathbf{p} = -\mu \mathbf{p} + \beta'(\mathbf{I}-\mathbf{P})\mathbf{A}\mathbf{p}
    \label{eq:qmf_vector}
\end{equation}
where $\mathbf{P}=\mathrm{diag}(p_1, \cdots, p_N)$. Near the absorbing-state threshold, the instability is controlled by the leading Perron mode of $\mathbf{A}$. We therefore denote the normalized principal eigenvector (PEV) by $\mathbf{u}$, with
\begin{equation}
    \mathbf{A}\mathbf{u} = \Lambda_1\mathbf{u}, \quad \mathbf{u}^{\rm T}\mathbf{u} = 1.
\end{equation}
Here $\Lambda_1$ is the largest eigenvalue of matrix $\mathbf{A}$. Projecting the Eq.\eqref{eq:qmf_vector} onto $\mathbf{u}$ gives
\begin{equation}
    \begin{aligned}
    \mathbf{u}^{\rm T}\Delta\mathbf{p}&=\mathbf{u}^{\rm T}[-\mu \mathbf{p} + \beta'(\mathbf{I}-\mathbf{P})\mathbf{A}\mathbf{p}]\\
    &=(\beta'\Lambda_1-\mu)\mathbf{u}^{\rm T}\mathbf{p}-\beta'\mathbf{u}^{\rm T}\mathbf{PAp}
    \end{aligned}
    \label{eq:projection}
\end{equation}
Close to the threshold, the critical eigenspace is spanned by the PEV, whereas the remaining modes are linearly stable and decay faster. We therefore write the probability vector, to leading order, as
\begin{equation}
    \mathbf{p} = x\mathbf{u}+O(x^2), \quad x\to0,
    \label{eq:main_mode}
\end{equation}
where $x$ is the amplitude of the PEV mode.This gives
\begin{equation}
\mathbf{u}^{\mathrm{T}}\Delta\mathbf{p} = \Delta x + O(x^2), \quad \mathbf{u}^{\mathrm{T}}\mathbf{p}
= x+O(x^2),
\label{eq:projection_terms_appendix}
\end{equation}
and
\begin{equation}
\rho = \frac{1}{N}\sum_i p_i = xM_1+O(x^2),
\label{eq:rho_mode_appendix}
\end{equation}
where
\begin{equation}
M_n = \frac{1}{N}\sum_{i=1}^{N}u_i^n .
\label{eq:moments_appendix}
\end{equation}
Using $\mathbf{Au}=\Lambda_1\mathbf{u}$, the nonlinear projection term becomes
\begin{equation}
\mathbf{u}^{\mathrm{T}}\mathbf{P}\mathbf{A}\mathbf{p}
= x^2\Lambda_1 N M_3 + O(x^3).
\label{eq:nonlinear_projection_appendix}
\end{equation}

Substituting Eqs.~\eqref{eq:projection_terms_appendix}--\eqref{eq:nonlinear_projection_appendix} into Eq.~\eqref{eq:projection} and keeping terms up to cubic order yields
\begin{equation}
    \Delta{x} = rx + cx^2 - dx^3
\end{equation}
with
\begin{equation}
    \begin{aligned}
        r &= \beta \Lambda_1-\mu,\\
        c &= \Lambda_1M_1(\alpha - \beta K),\\
        d &= \alpha \Lambda_1 M_1^2K.
    \end{aligned}
    \label{eq:rcd_appendix}
\end{equation}
Here $K=NM_3/M_1$. The coefficient $r$ determines the linear stability of the absorbing state. Therefore, the upper transition boundary is obtained from r=0, giving
\begin{equation}
    \beta_c^{\mathrm{high}} = \frac{\mu}{\Lambda_1}.
    \label{eq:appendix_beta_c_high}
\end{equation}
At this threshold, the linear term vanishes and the direction of the local branch is determined by the quadratic coefficient $c$. The reinforcement threshold separating the continuous and discontinuous regimes is obtained from $c=0$ at $\beta=\beta_c^{\rm high}$, yielding
\begin{equation}
    \alpha_c = \beta_c^{\mathrm{high}}K = \frac{\mu}{\Lambda_1}\frac{NM_3}{M_1}.
    \label{eq:appendix_alpha_c}
\end{equation}
For $\alpha<\alpha_c$, the active branch emerges continuously from the absorbing state. For $\alpha>\alpha_c$, the local branch bends backward, and the system develops bistability between the absorbing branch and a finite-prevalence active branch. In this regime, the active branch is created through a saddle-node bifurcation at a lower value of $\beta$. 

The nonzero stationary solutions of Eq.~\eqref{eq:amplitude_appendix} satisfy
\begin{equation}
    x_{\pm} = \frac{c\pm \sqrt{c^2+4dr}}{2d}.
    \label{eq:x_pm}
\end{equation}
Within the local amplitude description, $x_+$ corresponds to the stable active branch, whereas $x_-$ corresponds to the unstable saddle separating the basins of attraction. The saddle-node occurs when the discriminant vanishes,
\begin{equation}
    c^2+4dr=0.
    \label{eq:saddle_node_condition_appendix}
\end{equation}
Solving this condition gives the lower boundary
\begin{equation}
    \begin{aligned}
    \beta_c^{\mathrm{low}} &= \frac{2\sqrt{\alpha K\beta_c^{\mathrm{high}}}-\alpha}{K}\\
    &= \beta_c^{\mathrm{high}} - \frac{(\sqrt{\alpha} - \sqrt{\alpha_c})^2}{K}.
    \end{aligned}
    \label{eq:appendix_beta_c_low}
\end{equation}
The hysteresis width is therefore
\begin{equation}
    \Delta \beta = \beta_c^{\mathrm{high}} - \beta_c^{\mathrm{low}}= \frac{(\sqrt{\alpha} - \sqrt{\alpha_c})^2}{K}.
\end{equation}
This derivation shows that the discontinuous transition arises from the sign change of the quadratic coefficient in the PEV amplitude equation. Macro-level reinforcement contributes the positive term proportional to $\alpha$, while the susceptible-state factor contributes the negative saturation term proportional to $\beta K$. Their competition determines the reinforcement threshold $\alpha_c$, and the saddle-node condition determines the hysteresis width.

\section{Identifying the class of phase transition}\label{appendix:supp_transition_class}
To clarify the nature of the discontinuous transition in our model, we perform a finite-size scaling analysis following the framework for discontinuous nonequilibrium phase transitions into absorbing states~\cite{de2015generic}.

In finite stochastic systems with absorbing states, direct simulations eventually fall into the absorbing configuration. This makes it difficult to sample the long-time active dynamics near the transition. Quasistationary (QS) simulations avoid this problem by keeping the dynamics in the active subspace. The QS ensemble therefore allows us to measure the order-parameter distribution and its fluctuations in finite systems. In the thermodynamic limit, QS observables converge to the corresponding stationary quantities of the original process~\cite{dickman2002quasi}.

We focus on Erd\H{o}s--R\'enyi networks with $\avg{k}=10$ and different sizes, fixing the reinforcement strength at $\alpha=0.15$, where the transition is discontinuous. For each network size, we perform high-resolution QS sampling around the transition region and locate the finite-size pseudotransition point $\beta_\chi(N)$ from the peak of the susceptibility. The results show that the critical behavior belongs to the class of first-order transitions, characterized by coexistence between a low-activity phase and a high-activity phase.

\begin{figure*}
    \centering
    \includegraphics[width=\textwidth]{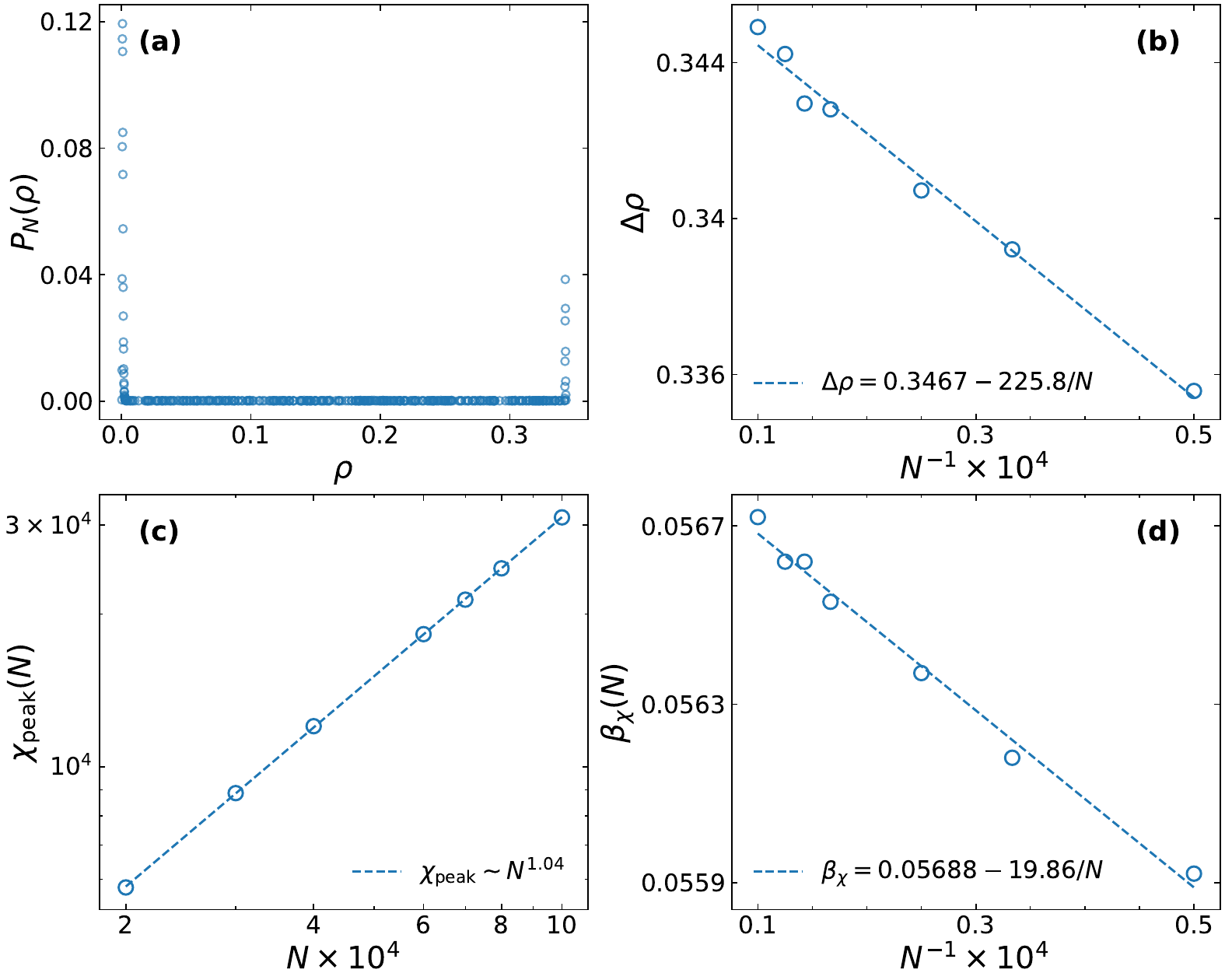}
    \caption{
    \textbf{Finite-size analysis of a first-order transition on ER networks.} All results are obtained at $\alpha=0.15$ on ER networks with $\avg{k}=10$. Panels (a)--(c) are evaluated at $\beta=\beta_{\chi}^c(N)$; panel(a) uses $N=5\times 10^4$.
    (a) Quasistationary prevalence distribution $P_N(\rho)$, showing two separated peaks associated with the low- and high-activity phases.
    (b) Peak separation $\Delta\rho(N)=\rho_+(N)-\rho_-(N)$ as a function of $1/N$. The linear extrapolation gives a finite value in the thermodynamic limit, indicating a nonzero order-parameter jump.
    (c) The susceptibility peak $\chi_{\rm peak}(N)$ grows approximately linearly with system size, consistent with size scaling expected for a first-order transition.
    (d) Finite-size pseudotransition point $\beta_\chi(N)$ as a function of $1/N$, showing an inverse-size finite-size shift. 
    }
    \label{fig:sfigure1}
\end{figure*}

For a first-order transition, the distribution of the prevalence can be written phenomenologically as the bimodal probability distribution
\begin{equation}
    P_N(\rho)
    \simeq
    W_-(N,\beta) P_N^{(-)}(\rho)
    +
    W_+(N,\beta) P_N^{(+)}(\rho),
    \label{eq:two_phase_distribution}
\end{equation}
where $P_N^{(-)}(\rho)$ and $P_N^{(+)}(\rho)$ are the distributions associated with the low- and high-activity phases, and $W_-$ and $W_+$ are their statistical weights. This two-peak structure is the finite-size signature of phase coexistence.

In the thermodynamic limit, a genuine first-order transition retains two distinct order-parameter values,
\begin{equation}
    \Delta \rho(N)
    =
    \rho_+(N)-\rho_-(N)
    \longrightarrow
    \Delta \rho_\infty >0,
    \label{eq:supp_jump_condition}
\end{equation}
where $\rho_-(N)$ and $\rho_+(N)$ denote the positions of the low- and high-activity peaks.
\begin{equation}
    \chi_N(\beta_\chi^c) \sim N
\end{equation}

The finite-size pseudotransition point $\beta_{\chi}(N)$ is estimated from the peak value of the susceptibility, which is defined as
\begin{equation}
    \chi_N(\beta) = N \frac{\avg{\rho^2} - \avg{\rho}^2}{\avg{\rho}}.
    \label{eq:supp_susceptibility}
\end{equation}
From the bimodal probability distribution, the variance of the prevalence contains an inter-phase contribution,
\begin{equation}
    \mathrm{Var}(\rho)
    \simeq
    W_-W_+(\rho_+-\rho_-)^2 + \mathrm{Var}_{\rm intra}(\rho)
\end{equation}
Thus, near the pseudotransition point
\begin{equation}
    \chi_N(\beta_\chi^c)
    \simeq
    \frac{N W_-W_+\Delta\rho^2}{\avg{\rho}} + \chi_{\rm intra}.
\end{equation}
The first term comes from the separation between the two coexisting phases, while $\chi_{\rm intra}$ denotes fluctuations inside each phase. For coexisting phases in first-order transition, the width of each peak decreases with system size, so that $\mathrm{Var}_\pm(\rho)\sim N^{-1}$, giving $\chi_{\rm intra}\sim O(1)$. Near coexistence, the two phases have comparable weight, $W_\pm =O(1)$. Since $\Delta \rho=O(1)$ and $\avg{\rho}=O(1)$ at a first-order transition, the dominant contribution scales as the system size $N$, giving

The same two-phase picture also explains the finite-size shift of the pseudotransition point. The relative phase weight has the generic form
\begin{equation}
    \frac{W_+}{W_-}
    \sim
    \exp \left[
    N C(\beta-\beta_c)
    \right],
\end{equation}
where $C$ is a nonzero constant. To observe coexistence in a finite system, the two weights must be comparable, so the exponent must remain of order unity. This gives
\begin{equation}
    N[\beta_\chi(N) - \beta_c(\infty)] \sim O(1),
\end{equation}
and therefore
\begin{equation}
    \beta_\chi(N) - \beta_c(\infty) \sim N^{-1}.
\end{equation}
Thus, $\beta_\chi(N)$ is expected to approach $\beta_c(\infty)$ with an inverse-size correction

These finite-size behaviors are tested in Fig.~\ref{fig:sfigure1}. Figure~\ref{fig:sfigure1}(a) shows a bimodal QS prevalence distribution at $\beta_\chi(N)$ for $N=5\times 10^4$, indicating coexistence between low- and high-activity phases. Figure~\ref{fig:sfigure1}(b) shows that the peak separation $\Delta\rho(N)$ remains finite under extrapolation to $N\to\infty$, confirming a nonzero order-parameter jump. Figure~\ref{fig:sfigure1}(c) shows that the susceptibility peak grows approximately linearly with system size, consistent with size scaling expected from two-phase coexistence. Figure~\ref{fig:sfigure1}(d) shows that $\beta_\chi(N)$ shifts approximately linearly with $1/N$, as predicted by inverse-size finite-size scaling.

Taken together, these results support a first-order transition governed by coexistence between low- and high-activity phases. The large fluctuations near the transition arise mainly from switching between the two coexisting phases, rather than from critical fluctuations within a single active phase.

\section{Numerical estimation of the reinforcement threshold $\alpha_c$}\label{appendix:numeric_alphac}
Following the use of an index to quantify the onset of bistability in hypergraph SIS dynamics~\cite{landry2020effect}, we define a branch-separation observable to identify the reinforcement threshold $\alpha_c$ in our work. For each reinforcement strength $\alpha$, we perform quasistationary simulations over a range of intrinsic transmission probabilities $\beta$, starting from low- and high-prevalence initial conditions. This gives two quasistationary branches,
\begin{equation}
\rho^{\rm L}(\alpha,\beta),
\quad
\rho^{\rm H}(\alpha,\beta),
\end{equation}
where $L$ and $H$ denote the low- and high-prevalence initializations, respectively.
\begin{figure*}
    \centering
    \includegraphics[width=\textwidth]{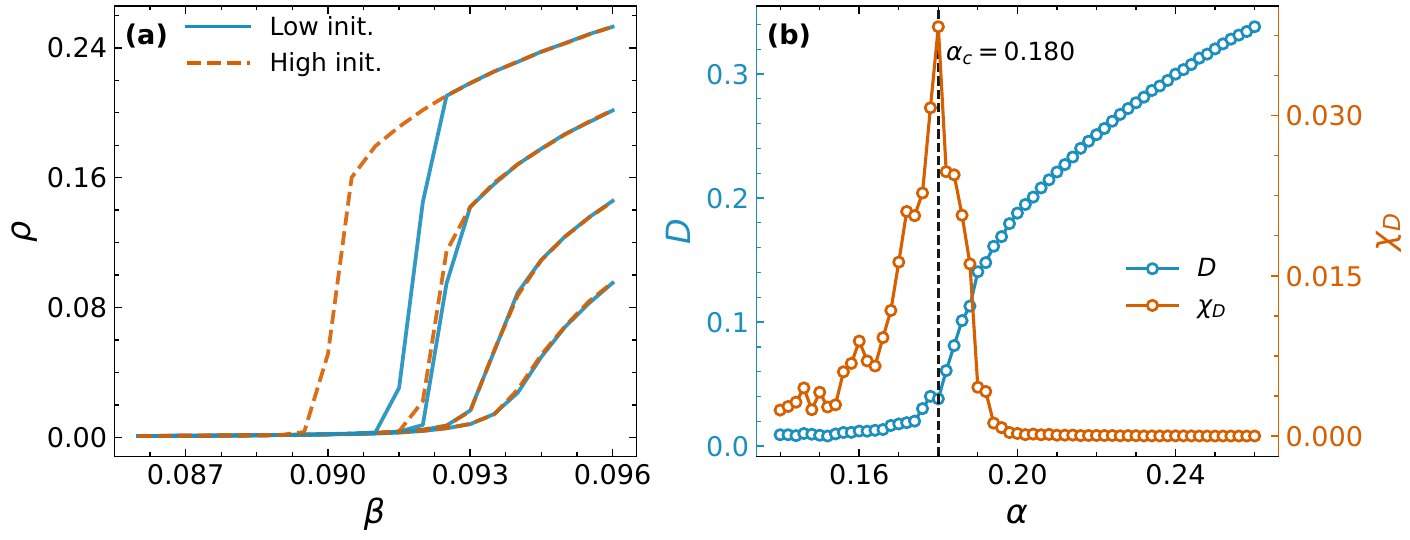}
    \caption{
    \textbf{Numerical estimation of the reinforcement threshold $\alpha_c$.}
    (a) Low- and high-prevalence quasistationary branches for several values of $\alpha$ in an ER network with $\avg{k}=6,N=10000,\mu=0.6$.
    (b) The bistability index $D(\alpha)$ and its susceptibility $\chi_D(\alpha)$.
    The dashed line and the red marker indicate the susceptibility peak, which defines
    $\alpha_c^{\rm num}=0.180$.
    }
    \label{fig:sfigure2}
\end{figure*}

Fig.~\ref{fig:sfigure2}(a) shows the corresponding low- and high-prevalence quasistationary branches for different values of $\alpha$. Below $\alpha_c^{\rm num}$, the two branches remain nearly indistinguishable. Near $\alpha_c^{\rm num}$, a finite separation starts to appear, and above $\alpha_c^{\rm num}$, a clear hysteresis window is formed. The resulting $\alpha_c^{\rm num}$ is used in the main text to compare the numerical onset of bistability with the theoretical prediction from the bifurcation analysis. We define a bistability index $D$ to characterize the onset of branch separations.
\begin{equation}
D(\alpha)=
\max_{\beta}
\left[
\rho^{\rm H}(\alpha,\beta)
-
\rho^{\rm L}(\alpha,\beta)
\right] ,
\end{equation}
The blue curve in Fig.~\ref{fig:sfigure2}(b) shows $\avg{D(\alpha)}$ as a function of $\alpha$. For small $\alpha$, the two branches nearly coincide and $D$ remains close to the finite-size fluctuation level. As $\alpha$ increases, $D$ grows, indicating the emergence of a separation between the two branches.

The numerical reinforcement threshold is determined from the susceptibility of $D$,
\begin{equation}
\chi_D(\alpha)
=
N
\frac{
\langle D^2(\alpha)\rangle
-
\langle D(\alpha)\rangle^2
}{
\langle D(\alpha)\rangle
}.
\end{equation}
Thus,
\begin{equation}
\alpha_c^{\rm num}
=
\arg\max_{\alpha}\chi_D(\alpha).
\end{equation}
The orange curve in Fig.~\ref{fig:sfigure2}(b) shows $\chi_D(\alpha)$, whose peak gives the corresponding $\alpha_c$. We use this procedure to obtain the numeric values of $\alpha_c$ reported in the main text.

\section*{Acknowledgements}
This work was supported by the Guangdong Basic and Applied Basic Research Foundation (Grant No.~2024A1515012692).

\section*{Author contributions}
G.Z. and P.-B.C. conceived the study. G.Z. performed the theoretical analysis and numerical simulations, analyzed the data, and wrote the original draft. P.-B.C. developed the model, supervised the project and contributed to the interpretation of the results. Both authors discussed the results, reviewed and edited the manuscript, and approved the final version.

\section*{Competing interests}
The authors declare no competing interests.

\section*{Data \& Code availability}
The custom data and code used in this study is available from the corresponding author upon reasonable request and will be deposited in a public repository upon publication.

\bibliography{apssamp}

\end{document}